\documentclass[%
 reprint,
 floatfix,
superscriptaddress,
 amsmath,amssymb,
 aps,
 prx,
]{revtex4-2}

\usepackage{graphicx}
\usepackage{dcolumn}
\usepackage{bm}
\usepackage{svg}
\usepackage{microtype} 
\usepackage{paralist}
\usepackage[normalem]{ulem}
\usepackage[unicode=true, colorlinks=true, citecolor={blue!80!black}, urlcolor={blue!50!black}, linkcolor = {blue!80!black}]{hyperref} 
\usepackage{array}
\usepackage{siunitx}
\usepackage{wrapfig}
\usepackage{enumitem}
\usepackage{amsmath,amssymb}
\usepackage{array}
\usepackage{soul}
\usepackage{centernot}
\usepackage{bibunits}
\usepackage[titletoc, title]{appendix}
\usepackage{titletoc}
\usepackage{lineno}
\usepackage{lipsum}  
\usepackage{mathtools}
\usepackage{multirow}
\usepackage{graphicx}
\begin{document}

\title{
Fabrication effects on Niobium oxidation and surface contamination in Niobium-metal bilayers using X-ray photoelectron spectroscopy
}

\author{Tathagata Banerjee}
\email{tb548@cornell.edu}
\affiliation{School of Applied and Engineering Physics, Cornell University, Ithaca, NY 14853, USA}

\author{Maciej W. Olszewski}
\affiliation{Department of Physics, Cornell University, Ithaca, NY 14853, USA}

\author{Valla Fatemi}
\email{vf82@cornell.edu}
\affiliation{School of Applied and Engineering Physics, Cornell University, Ithaca, NY 14853, USA}

\date{\today}

\begin{abstract}
Superconducting resonators and qubits are limited by dielectric losses from surface oxides. 
Surface oxides are mitigated through various strategies such as the addition of a metal capping layer, surface passivation, and acid processing.
In this study, we demonstrate the use of X-ray photoelectron spectroscopy (XPS) as a rapid characterization tool to study the effectiveness cap layers for niobium for further device fabrication.
We non-destructively evaluate 17 capping layers to characterize their ability to prevent oxygen diffusion, and the effects of standard fabrication processes -- annealing, resist stripping, and acid cleaning.
We downselect for resilient capping layers and test their microwave resonator performance. 

\end{abstract}

\maketitle

\section{Introduction}

Superconducting qubit devices have seen rapid advancement in performance metrics with integration of improved materials and fabrication procedures~\cite{mcrae_materials_2020,murray_material_2021}.
Recent results demonstrate that conventional transmons are a promising candidate for quantum computing~\cite{bal_systematic_2024,tuokkola_methods_2025,bland_millisecond_2025,chang_eliminating_2025}, with the surface oxide of the superconducting metals such as niobium (Nb), tantalum (Ta), and aluminum (Al) as the suspected source as a major source of loss~\cite{burnett_analysis_2016,de_leon_materials_2021,crowley_disentangling_2023,murthy_identifying_2025,krasnikova_experimental_2025}.
Nb, in particular, is chemically resilient, has a high critical temperature, and grows in superconducting structure on many substrates; however it has a non-self-limiting lossy oxide~\cite{murthy_identifying_2025,murthy_developing_2022,romanenko_three-dimensional_2020,bafia_oxygen_2024,wenskat_vacancy_2022,altoe_localization_2022}.
Different methods of mitigating the oxide have been reported, including encapsulation~\cite{bal_systematic_2024,de_ory_low_2025,zhou_ultrathin_2024,karuppannan_improved_2024}, surface passivation~\cite{alghadeer_surface_2023,zheng_nitrogen_2022}, and acid treatments~\cite{verjauw_investigation_2021,gingras_improving_2025}.
To this end, superconducting resonator losses serve as a rapid turnaround characterization technique for determining the performance potential of materials for qubit applications~\cite{mcrae_materials_2020}, while additional materials characterization techniques are required for understanding the origin of the loss.

\begin{figure}[t]
    \centering
    \includegraphics[width=0.9\linewidth]{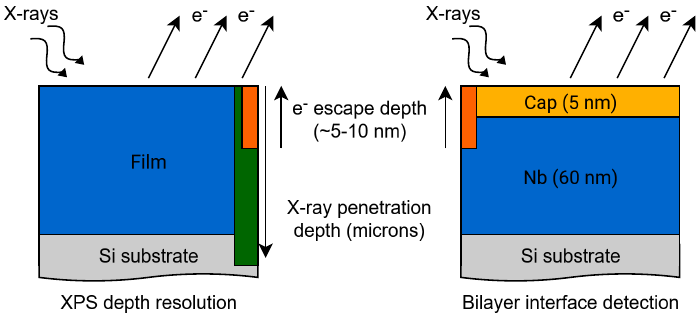}
    \caption{
    Schematic of X-rays entering the measured sample.
    Scattered electrons escape from 5-10~\si{\nano\meter} of depth.
    XPS scan of \SI{5}{\nano\meter} film will capture electrons from both the metal cap and Nb.
    }
    \label{fig:schematic}
\end{figure}

In this study, we characterize the effect of multiple standard processes conducted during fabrication -- annealing, resist stripping, and acid cleaning -- on superconducting bilayers composed of niobium with a cap layer.
We disentangle the various effects by measuring the oxidation state of Nb with X-ray photoelectron spectroscopy (XPS), and leverage the depth sensitivity of the XPS, as shown in Figure~\ref{fig:schematic}.
While the X-rays penetrate several microns into the sample, the electrons escape from the top 5-10~\si{\nano\meter}, as determined by the inelastic mean free path of electrons in various materials~\cite{powell_practical_2020}. 
By having a thin encapsulation layer, we can non-destructively measure the Nb-metal interface and determine the amount of oxidized Nb.

We report on the efficacy of various metals, nitrides, and alloys as oxygen diffusion barriers, both as deposited and after annealing. 
We examine the surface contamination and damage after processing samples with an Integrated Micro Materials AZ 300T chemical resist strip bath~\cite{olszewski_low-loss_2025}.
We also subject the Nb-metal bilayers to various acid cleaning processes, using Nanostrip, hydrofluoric acid (HF), and buffered oxide etch (BOE)~\cite{place_new_2021,torrescastanedo_formation_2024}. 
Finally, we fabricate resonators using a subset of the Nb-metal bilayers to examine the effect of the best caps on the resonator performance. 

\section{Materials and methods}

\subsection{Sample preparation}

Samples were sputtered in an AJA International, Inc. magnetron sputtering chamber, as used in Olszewski et al.~\cite{olszewski_low-loss_2025}. 
We deposited \SI{60}{\nano\meter} Nb films on intrinsic high-resistivity ($\geq10,000$ $\Omega$-cm) float zone silicon (100) \SI{100}{\milli\meter} wafers followed by in-situ sputtering of the \SI{5}{\nano\meter} capping layer. 
We analyze metals -- Al, Au, Hf, Mo, Pd, Pt, Ta, Ti, W, and Zr; nitrides -- NbN, ZrN, HfN, TiN, and TaN; and alloys -- Ti-W and Zr-Y.

Full details of the sample preparation is provided in Appendix~\ref{app:prep}.
The critical temperatures (T\textsubscript{c}) of the samples were measured to ensure that the T\textsubscript{c} had not been reduced significantly from that of Nb, \SI{9.24}{\kelvin} (Appendix \ref{app:Tc}).

\subsection{XPS sample processing}

Samples were subjected to three different processes: \textbf{annealing}, \textbf{strip bath}, or \textbf{acid cleaning}, conducted on the day of the XPS measurements.
All XPS measurements were conducted in a ThermoScientific Nexsa G2 Surface Analysis system with a chamber pressure around 2 $\times$ 10$^{-7}$~\si{Torr}. Further system details are provided in Appendix~\ref{app:XPS}.
The measurements were conducted on un-etched surfaces of the bilayer films to examine the amount of oxide in the Nb. 

We tested the robustness of the capping layers with two different \textbf{annealing} tests: air and vacuum.
Annealing is a common practice in device fabrication, and high temperatures are used for hydrocarbon removal.
We conducted both experiments at \SI{200}{\celsius} for \SI{1}{\hour} with air-annealing done on a hot plate in the fume hood, and vacuum-annealing done in a tube furnace with pressure below $<5\times10^{-6}$~\si{Torr}.
The proclivity of capping layers to allow Nb oxidation at \SI{200}{\celsius} is likely indicative of trace oxidation taking place at lower temperatures.
This may not be easily detectable at lower thermal exposures due to the resolution of the XPS, and yet still be a source of loss.

For the \textbf{strip bath} test, samples were first spin coated with Microposit S1813 resist and baked at \SI{90}{\celsius} for \SI{1}{\minute}. Samples were then soaked in a 80-90~\si{\celsius} heated bath of as-received AZ 300T stripper for one hour.
The bath was followed by sonication in isopropanol and deionized (DI) water, and rinses in DI water and isopropanol. 
Samples were blow-dried with ultra-high purity nitrogen. 

For \textbf{acid cleaning} tests, the samples were dipped in three acids: hydrofluoric acid (HF), buffered oxide etch (BOE), and Nanostrip. 
For HF, a 2\% HF dip for \SI{1}{\minute} was followed by rinsing in DI water and blow-drying.
The BOE dip was done in a 10:1 BOE solution for \SI{5}{\minute}, followed by rinsing in DI water and blow-drying.
The Nanostrip clean was done in a \SI{30}{\minute} soak in \SI{70}{\celsius} Nanostrip solution (a mixture of sulfuric acid, peroxymonosulfuric acid, and hydrogen peroxide), followed by rinsing in DI water and blow-drying. 

\subsection{Resonator measurements}

The fabrication process of the resonators is detailed in Appendix \ref{app:resonator}, and follows the fabrication in \cite{olszewski_low-loss_2025}.
Resonators were fabricated using a design from the open-source Boulder Cryogenic Quantum Testbed containing eight branches per sample with target frequency of 4-8 \si{\giga\hertz}~\cite{kopas_simple_2022}. 
Samples were measured in a Bluefors small diameter cryostat with a base temperature of \SI{500}{\milli\kelvin}.
For each resonance, we vary the probe frequency and fit the complex-valued transfer function to the expected resonance response.
We extract the medium power loss, $\delta_{\rm MP}$, for an average photon number $\langle n \rangle <100$ and high power loss  $\delta_{\rm HP}$ ($\langle n \rangle >10^6$) by fitting the experimentally measured internal quality factor~\cite{khalil_analysis_2012}.
The lower power measurements are restricted by a higher noise photon occupation number of the \SI{500}{\milli\kelvin} cryostat, which is on the order of a 1-10 photons (Appendix~\ref{app:resonator_measurements}).
Thus, $\delta_{\rm MP}$ loss is extracted from data collected with $\langle n \rangle$ between $10-100$. 
We do not fit the data for two-level-system (TLS) loss $\delta_{\rm TLS}$ as the measurements do not show saturation at high power or medium power~\cite{crowley_disentangling_2023,verjauw_investigation_2021}.
The fridge setup and measurement schematics are detailed in Appendix \ref{app:resonator}.

\section{Results}

\subsection{Baseline and annealing}
\begin{figure}[]
    \centering
    \includegraphics[width=1\linewidth]{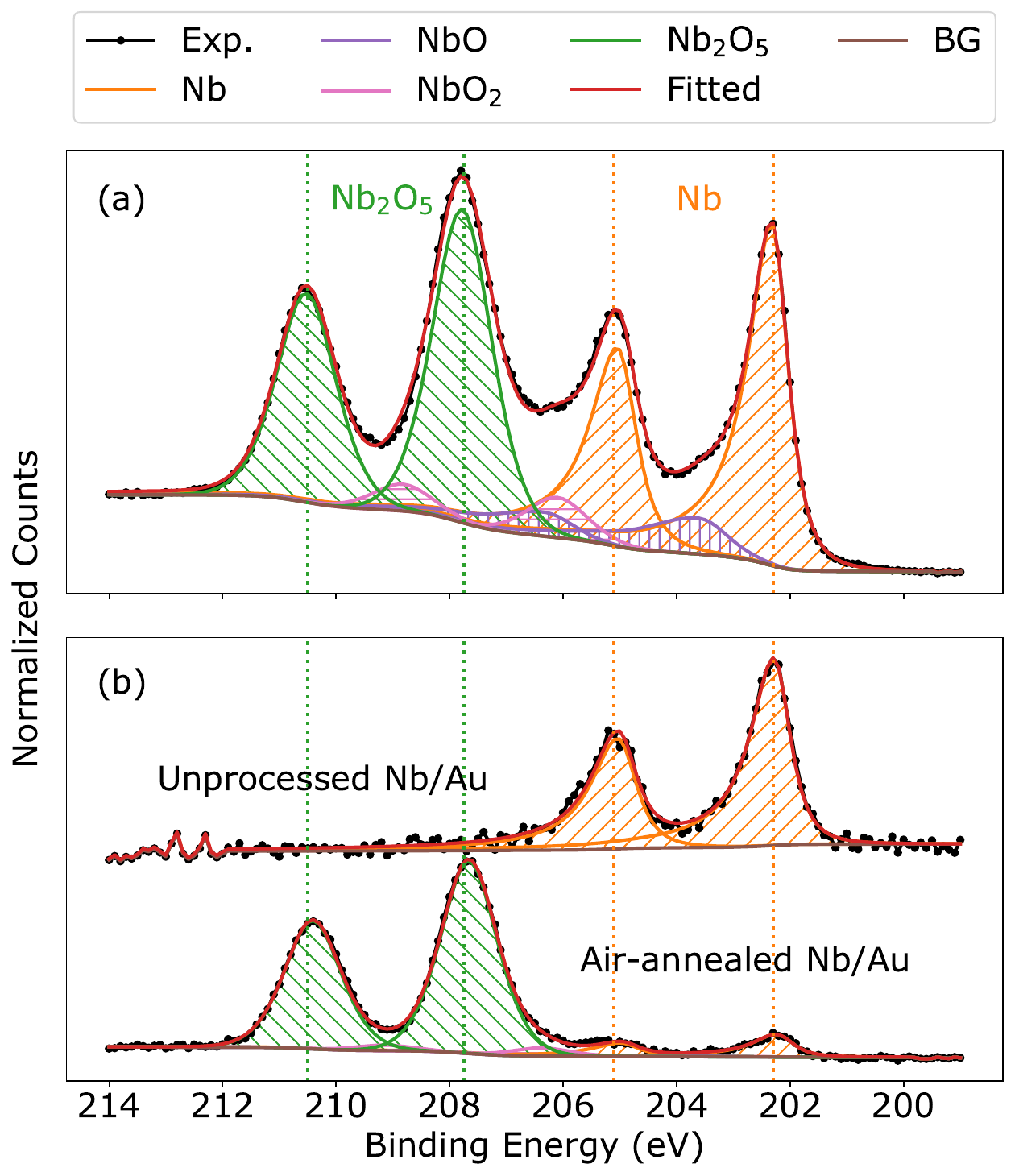}
    \caption{Nb 3d core level spectrum of (a) uncapped niobium and (b) unprocessed and air-annealed Nb/Au. 
    }
    \label{fig:Nb_pure}
\end{figure}

We first discuss the results for the uncapped Nb sample, shown in Figure \ref{fig:Nb_pure}(a).
We fit the niobium metal and the NbO peaks with asymmetric Voigt-like lineshapes, as both are conducting.
The other oxides are fit with symmetric Gaussian-Lorentzian peaks.
We determine the presence of Nb oxides based on the Nb peaks from the fit. 
We see a significant amount of Nb\textsubscript{2}O\textsubscript{5} on the Nb surface, with smaller amounts of other suboxides.

By in-situ sputtering Nb and metal layers consecutively, we remove the possibility of naturally oxidizing the Nb surface prior to extraction from the deposition chamber.
The samples are loaded into the XPS without additional processing.
Therefore, any oxidation seen of the Nb will be through diffusion of oxygen through the surface encapsulation layer.
We use this to determine which metal layers are effective in preventing oxidation during the various fabrication steps, and down-select the most promising candidates.

\begin{figure}[]
    \centering
    \includegraphics[width=1\linewidth]{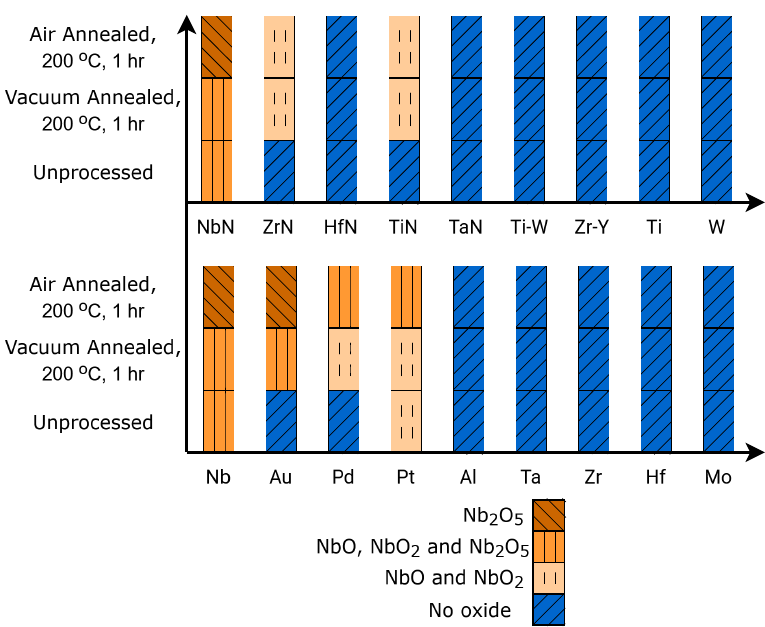}
    \caption{
    The relative oxidation of the Nb for the unprocessed films, after annealing in air, and after vacuum annealing.
    Un-oxidized Nb is indicated in blue.
    Increasing shade of orange indicate presence of higher oxidation states.
    Note that higher shades do not indicate higher relative percentages of the oxides.}
    \label{fig:annealing_table}
\end{figure}

Identical Nb fits were applied for capped samples.
In Figure \ref{fig:Nb_pure}(b), we compare the unprocessed and air-annealed Au-capped film.
There is no oxide under the Au cap in the Nb layer before processing.
After annealing, however, oxygen has diffused through the capping layer and the Nb is significantly oxidized.
We summarize the results of all samples in Figure~\ref{fig:annealing_table}, where we categorize the bilayers based on the amount of Nb oxide.
Detailed XPS spectra of each bilayer are in Appendix~\ref{app:anneal}.

The Al, Hf, Mo, Ta, Ti, W, and Zr-capped films, as well as the alloy-capped films do not form Nb oxide under the capping layer.
The Zr-capped sample has a fully oxidized Zr layer, without any Nb oxidation beneath it.
This motivated the use of ZrO\textsubscript{x} as a tunnel barrier to fabricate all refractory metal Nb/ZrO\textsubscript{x}/Nb trilayer Josephson junctions \cite{choi_low_2025}, which is more chemically resilient than traditional liftoff AlO\textsubscript{x}-based junctions~\cite{dolan_offset_1977,quintana_characterization_2014,muthusubramanian_wafer-scale_2024}.

The noble metals, namely Au, Pd, and Pt, perform poorly as an oxygen diffusion barrier.
These do not prevent oxygen dissolving or otherwise traveling through the grain boundaries to oxidize the Nb at the interface, as the caps themselves do not oxidize.
Au, Pd, and Pt capped films also have an increase in the Nb oxidation after annealing in air, which promotes the transfer of oxygen into the interface. 
This may be a result of having thinner capping films, and thicker films may prevent the oxidation of the Nb underneath.

Nitrides are promising for capping layers, as they have thinner oxide layer, as the nitrogen occupies vacancies and interstitials~\cite{logothetidis_room_1999,lubenchenko_native_2019,zier_xps_2005}. 
The XPS fitting is made impossible by the appearance of a NbN\textsubscript{x} peak, which may have been formed when the Ar and N\textsubscript{2} gas mixture is first introduced into the chamber, prior to the reactive sputtering~\cite{carvalho_electronic_2013,dickey_interaction_1975}.
As a result, we are unable to resolve the nitride and suboxide presence due to the overlap of the NbN\textsubscript{x} peak with the NbO peak, noting that nitride samples marked as `no oxide' are uncertain.
We do see a clear oxide peak increase for TiN and ZrN.

For NbN, it is not possible to verify the origin of the oxide, either the NbN cap or the Nb base layer, without angle-resolved spectroscopy. 
The significant amount of oxide observed in the NbN-capped sample made it a non-promising candidate.

We compare the spectra post-annealing to the unprocessed samples. 
A new peak is considered to be a formation of a suboxide, as Nb will likely not form a nitride at \SI{200}{\celsius}~\cite{semione_niobium_2019}.
HfN and TaN have no new peak formation after annealing, while ZrN and TiN have a minor increase (see Appendix \ref{app:anneal}).

The performance of the unprocessed samples is also understood through a theoretical model from Chaudhari et al.~\cite{chaudhari_active-learning_2025} that classifies metals as effective diffusion barriers based on DFT-calculated oxygen vacancy and interstitial defect energies.
Their study uses the data in Figure~\ref{fig:annealing_table} to iteratively train and validate the model within a feedback loop for selecting additional metals to test experimentally.
The details of this comparison are given in that work.

\subsection{Chemical strip}

We examine the surface contamination and damage after processing samples with an AZ 300T chemical strip bath, an N-methylpyrrolidone (NMP) based stripper with tetramethylammonium hydroxide (TMAH) additive~\cite{az_electronic_materials_az_2015}.
Resist residues and surface preparation are well known to affect performance of devices, either due to structural or contamination related losses~\cite{murthy_tof-sims_2022,quintana_characterization_2014,crowley_disentangling_2023}.
In Olszewski et al.~\cite{olszewski_low-loss_2025}, we found that despite the increase in the Nb oxide due to the AZ 300T strip bath, the performance was improved compared to other chemical strips, with the primary difference being the reduced presence of carbon, chlorine and nitrogen, contamination.
Evaluating the resilience of the bilayers to a chemical bath of AZ 300T will, therefore, assist with integration of the capping layer with existing fabrication processes.

Additional processing, such as etching and descum, do alter the resist and affect the contamination, even where the surface is not exposed, due to resist hardening or reaction with the plasma environment.
Our study provides information regarding the baseline process effect on the surface contamination and damage, and detailed studies should be conducted to characterize effects of additional processing for those that use them. 
The contaminants are summarized in Table~\ref{tab:AZ 300T} along with the relative atomic percentage of the carbon, with detailed survey spectra available in Appendix~\ref{app:AZ}.

We find that Mo, W, and Ti-W films were eroded by the strip process and the niobium was significantly oxidized (marked with stars in Table \ref{tab:AZ 300T}), making these films incompatible with heated AZ 300T processing.

\begin{table}[]
    \centering
    \begin{tabular}{c|c|c|c|c|c|c} \hline 
         Cap&  C At\% &N&Na&Ca &Si &Others\\ \hline \hline 
         --& 10.3\%& & \checkmark& \checkmark& &\\ \hline 
 Al& 13.2\%& & &\checkmark & &S\\ \hline 
 Au& 26.3\%& & & & &\\ \hline 
 Hf& 13.7\%& & \checkmark& & &F\\ \hline 
 Mo*& 16.6\%& & \checkmark& \checkmark&\checkmark &\\ \hline 
 Pd& 39.8\%& & & &&\\ \hline 
 Pt& 21.4\%& \checkmark& \checkmark& & &\\ \hline 
 Ta& 19.6\%& & \checkmark& \checkmark&\checkmark&S\\ \hline 
 Ti& 15.4\%& \checkmark& & & &\\ \hline 
 W*& 19.6\%& & \checkmark& \checkmark& &\\ \hline 
 Zr& 13.6\%& & & &\checkmark &\\ \hline \hline
 Ti-W*& 16.2\%& & \checkmark& \checkmark& \checkmark&\\ \hline
 Zr-Y& 14.1\%& & & &\checkmark &\\ \hline \hline
 NbN& 14.5\%& \checkmark& & \checkmark&\checkmark &\\ \hline 
 HfN& 16.9\%& \checkmark& \checkmark& & &F\\ \hline
 TaN& 17.3\%& \checkmark& & \checkmark&\checkmark &\\ \hline
 TiN& 14.6\%& \checkmark& & \checkmark& \checkmark&\\ \hline
 ZrN& 16.0\%& \checkmark& & &\checkmark &\\ \hline

 \end{tabular}
    \caption{Contamination of capped metal surfaces post-AZ 300T strip baths. 
    We compare the relative atomic percentage of carbon, and other contaminants, mainly N, Na, Ca, and Si.
    Mo, W, and Ti-W (marked with a $*$) were etched during the strip bath, and the Nb beneath the caps was oxidized.}
    \label{tab:AZ 300T}
\end{table}

Besides C; Na, Ca, N, and Si are the primary contaminants.
The uncapped Nb film were contaminated with Na and Ca on the surface, likely originating from the strip bath.
Si possibly originates from re-deposition or Si dust from the substrate.
It is important to note here that the presence of the contaminant is likely stochastic, and the trace amounts may vary upon repetition, as contamination is not uniform on the sample surface.
The XPS has a detection limit of around 0.1\%, thus there is a greater uncertainty in the exact percentage of the trace contamination.
The resulting table is indicative of the proclivity to physisorb contaminants rather than the exact contaminants themselves, i.e. samples that do not have Ca may have Ca after repetition of the experiment. 

Au-, Pd-, Zr-, and Zr-Y-capped films have the fewest contaminants on the surface, indicative of high potential for improving resonator performance without any acid cleaning prior to measurement. 
The F contamination in both Hf- and HfN-capped samples are unresolved, as there is no F-based processing.
One possible explanation is light contamination of the XPS chamber, however, measurements of standards do not show any fluorine on other surfaces. 

\subsection{Acid cleaning}

HF processing removes surface oxides and improves resonator and qubit performance~\cite{verjauw_investigation_2021,gingras_improving_2025}. 
The impact of the acids on the bilayers, for oxygen diffusion and reactivity, is important for selecting good bilayers.

\begin{figure}[]
    \centering
    \includegraphics[width=1\linewidth]{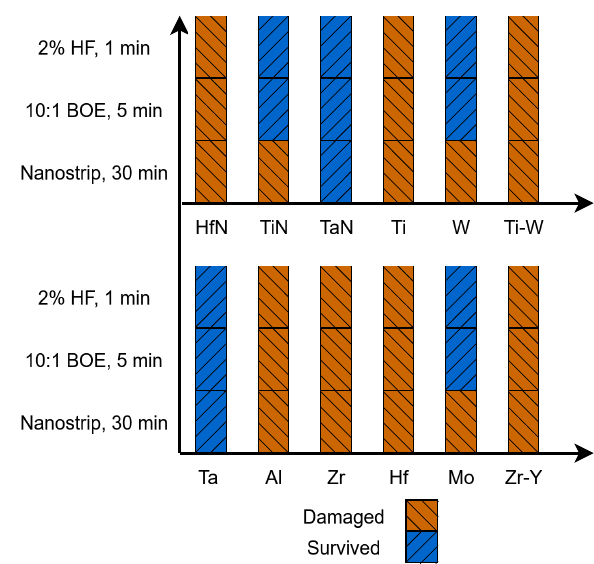}
    \caption{Results of acid cleaning of Nb-metal bilayers.
    2\% HF, 10:1 BOE, and Nanostrip acid treatments are tested.
    Samples that survived are marked in blue, while caps that were damaged are marked in orange.
    We define `damaged' as caps which were etched with the Nb film oxidized.}
    \label{fig:acid_table}
\end{figure}

The results of the acid cleaning are summarized in Figure \ref{fig:acid_table}. 
We define `damaged' as layers which are etched with oxidized Nb visible in the spectrum.
We only test caps that have do not have any Nb oxide under the cap formed during annealing (i.e. all except Au, Pd, Pt, ZrN and TiN). 

Al, Hf, Ti, Zr, HfN, and Zr-Y are significantly damaged by all three processes (2\% HF, 10:1 BOE, and Nanostrip), likely due to the high reactivity of these layers to the acids.
We cannot, however, exclude the effects poor wetting of these films on Nb at this time.
Ti-W is also damaged by all three processes, where Ti is completely removed, but W remains, and the Nb is not oxidized beneath. 
Mo, TiN, and W are etched by the Nanostrip, exposing the underlying Nb, but survive the HF and BOE. 
Mo, TiN, and W are promising for the use as caps for resonators, if used in conjunction with a resist stripper that does not etch the Mo or W.

Finally, TaN and Ta are not affected by any of the acid cleans. 
Ta and TaN are resilient to acid processing, and used in resonator and qubit fabrication~\cite{chang_eliminating_2025}.
Further details of all acid cleans are provided in Appendix \ref{app:Acid}.

\subsection{Resonator measurements}
 
\begin{figure}[]
    \centering
    \includegraphics[width=1\linewidth]{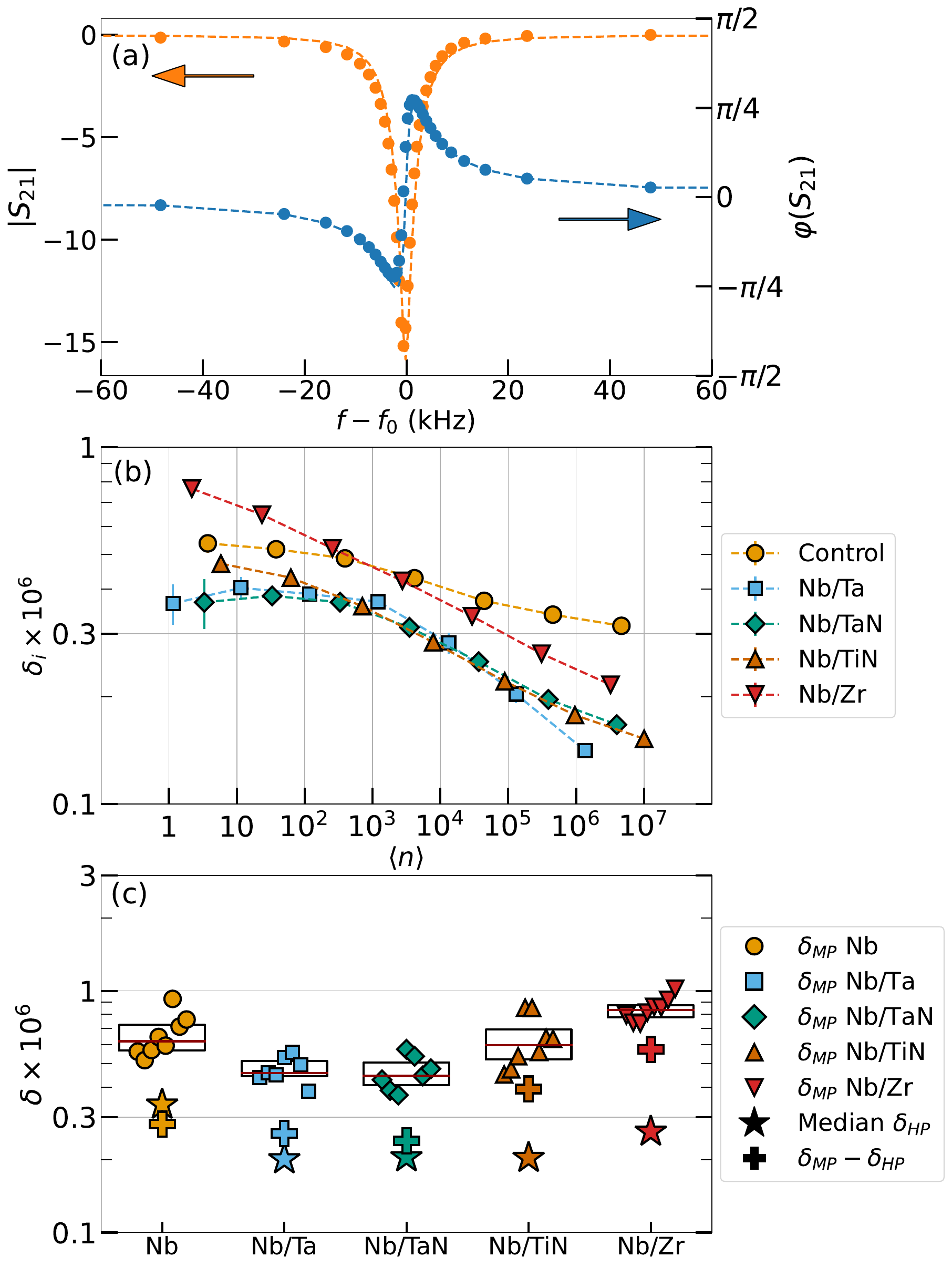}
    \caption{Resonator measurements a Nb resonator and capped Nb resonators. 
    (a) Sample Nb resonator measurement at $\langle n \rangle \approx 4.6\times 10^6$ photons at \SI{500}{\milli\kelvin}. 
    Magnitude (orange) and phase (blue) of $S_{21}$ is indicated with points, with fit shown in dashed lines.
    (b) Sample $\delta_i$ vs. $\langle n \rangle$ curves for the measured samples. 
    (c) Box plot of the medium power loss $\delta_{\rm MP}$ of the different resonators.
    Red line indicates the median $\delta_{\rm MP}$.
    Star indicates the median high power loss $\delta_{\rm HP}$, while the plus indicates the difference in median $\delta_{\rm MP}$ and $\delta_{\rm HP}$. 
    }
    \label{fig:resonators}
\end{figure}

We selected Ta, Zr, TaN, and TiN, as promising capping layers and fabricated resonators using these films.
A sample fit is shown of the Nb control resonator in Figure~\ref{fig:resonators}(a). 
Loss $\delta_i$ vs photon number $\langle n \rangle$ is shown in Fig.~\ref{fig:resonators}(b) for the respective capped films.
Fig.~\ref{fig:resonators}(c) shows resonator data at high-power ($\delta_{\rm HP}$) and medium-power ($\delta_{\rm MP}$) of Nb and select capped Nb films without acid cleaning prior to the measurement.
The Ta and TaN-capped samples have a lower $\delta_{\rm MP}$ than the control sample, highlighting the effectiveness of the cap. 
The TiN sample has a comparable $\delta_{\rm MP}$ to the Nb sample, however it has a lower median $\delta_{\rm HP}$, comparable to Ta and TaN samples.
This suggests higher power-dependent losses, and consequently a higher density of GHz-frequency two-level systems in the TiN-capped sample compared to the other two. This can be confirmed in future measurements at dilution refrigerator temperatures. 

Finally, the Zr-capped samples exhibit higher loss than the control sample, although similar to TiN, has a better median $\delta_{\rm HP}$ compared to the control. 
The Zr is fully oxidized in the sample, and thus the losses from the thicker Zr oxide may also be significant and comparable to the loss from the Nb oxide in the uncapped sample.

XPS measurements of samples post-fabrication (Table \ref{tab:SI-ResXPS} and Figure \ref{fig:SI-ResXPS}) showed comparable surface contamination, likely indicating that the difference in performance seen is correlated to the relative quality of the surface oxide of the different capping layers. 
The oxide's thickness, loss tangent, and dielectric constant all contribute to the resonator performance.
The intrinsic loss tangents can be lower and dielectric constants can be higher for the respective oxides from the caps tested compared to that of niobium oxide to improve performance~\cite{wenner_surface_2011}.
However, an increased thickness of the oxide, like in Zr, can increase losses to counteract the benefit from the alternative oxide.

\section{Conclusions}

We utilized non-destructive tests to evaluate Nb-metal bilayers to examine the oxygen diffusion, and the resilience of the caps to various standard processing steps. 
We have successfully down-selected certain capping layers that could be used for niobium capping to prevent oxygen diffusion. 
We show that \SI{5}{\nano\meter} thick noble metals are not suitable as capping layers, and see that annealing may promote oxidation in samples which otherwise do not have any at the interface.
Capped samples can react during both resist stripping and acid cleaning, limiting the possible list of materials for high-quality resonator. 
The data can be used to make decisions about materials to use for niobium fabrication of superconducting quantum devices.

\section*{Data and code availability}
The data that support the findings of this study are openly available in Zenodo at http://doi.org/10.5281/zenodo.17665093 \cite{banerjee_dataset_2026}.

\section*{Acknowledgments}
The authors thank J.T. Paustian, Jaehong Choi, Sarvesh Chaudhari, Tomas Arias, and Gregory Fuchs for discussions and feedback. 
We thank Kushagra Aggarwal and Haoran Lu for assistance with cryogenic setup and microwave measurements. 
We thank Alicia Tripp from the Cornell Center for Materials Research for advice regarding XPS measurements and fitting.

This material is based upon work supported by the Air Force Office of Scientific Research under award number FA9550-23-1-0706. Any opinions, findings, and conclusions or recommendations expressed in this material are those of the author(s) and do not necessarily reflect the views of the United States Air Force.

This work made use of the Cornell Center for Materials Research shared instrumentation facility. 
This work was performed in part at the Cornell NanoScale Facility, a member of the National Nanotechnology Coordinated Infrastructure (NNCI), which is supported by the National Science Foundation (Grant NNCI-2025233).

\section*{Author Contributions}

MO deposited all the films and conducted T\textsubscript{c} measurements. MO fabricated the resonators with help from TB. TB conducted resonator measurements and analysis with help from MO. TB conducted XPS measurements, subsequent fitting, and analysis of results. TB wrote the manuscript, which was edited by MO and VF. VF supervised the research. 

\section*{Competing Interests}
The authors have no conflicts to disclose.

\clearpage

\onecolumngrid
\begin{center}
    \textbf{\large SUPPLEMENTARY MATERIAL: Fabrication effects on Niobium oxidation and surface contamination in Niobium-metal bilayers using X-ray photoelectron spectroscopy} \\
    \vspace{10pt}
    Tathagata Banerjee,$^{1}$ Maciej W. Olszewski,$^{2}$ and Valla Fatemi$^{1}$ \\
    \vspace{2pt}
    \small{$^{1}$\textit{School of Applied and Engineering Physics, Cornell University, Ithaca, NY 14853, USA}} \\
    \small{$^{2}$\textit{Department of Physics, Cornell University, Ithaca, NY 14853, USA}}
\end{center}
\vspace{20pt}
\twocolumngrid

\appendix
\renewcommand{\thefigure}{S\arabic{figure}}
\renewcommand{\thetable}{S\arabic{table}}

\setcounter{figure}{0}
\setcounter{table}{0}

\section{Sample preparation}\label{app:prep}

Sample preparation follows \cite{olszewski_low-loss_2025} and we reiterate the preparation here for completeness.

Films for resonators were deposited on intrinsic high resistivity ($>10,000$ $\Omega$-cm) undoped float zone four-inch Si (100) wafers.
All elemental films -- Al, Au, Hf, Mo, Pd, Pt, Ti, W, and Zr -- as well the alloy Zr-Y were sputtered using a 2" diameter, and either a 1/4" or 1/8" thick target.
Depositions were conducted with ultra high-purity Ar gas from Airgas, a target-to-substrate distance of 20-30~\si{\centi\meter}, and substrate rotation at \SI{50}{rpm}. 
Nitride films were sputtered by flowing Ar and N\textsubscript{2} gas into the chamber during sputtering, with the ratios determined on a case-by-case basis for each nitride.
The Ti-W films was formed by co-sputtering from two targets (identical to the ones used for the respective elemental films) at approximately the same rate and checked in XPS.
The deposition rate was under \SI{1}{\nano\meter}/s for all targets with powers under \SI{400}{\watt}. 
The base pressure of the system varied between $1\times 10^{-9}$ and $5\times 10^{-8}$~\si{Torr}.

\section{Critical temperature measurements}\label{app:Tc}

\begin{figure*}[]
    \centering
    \includegraphics[width=\linewidth]{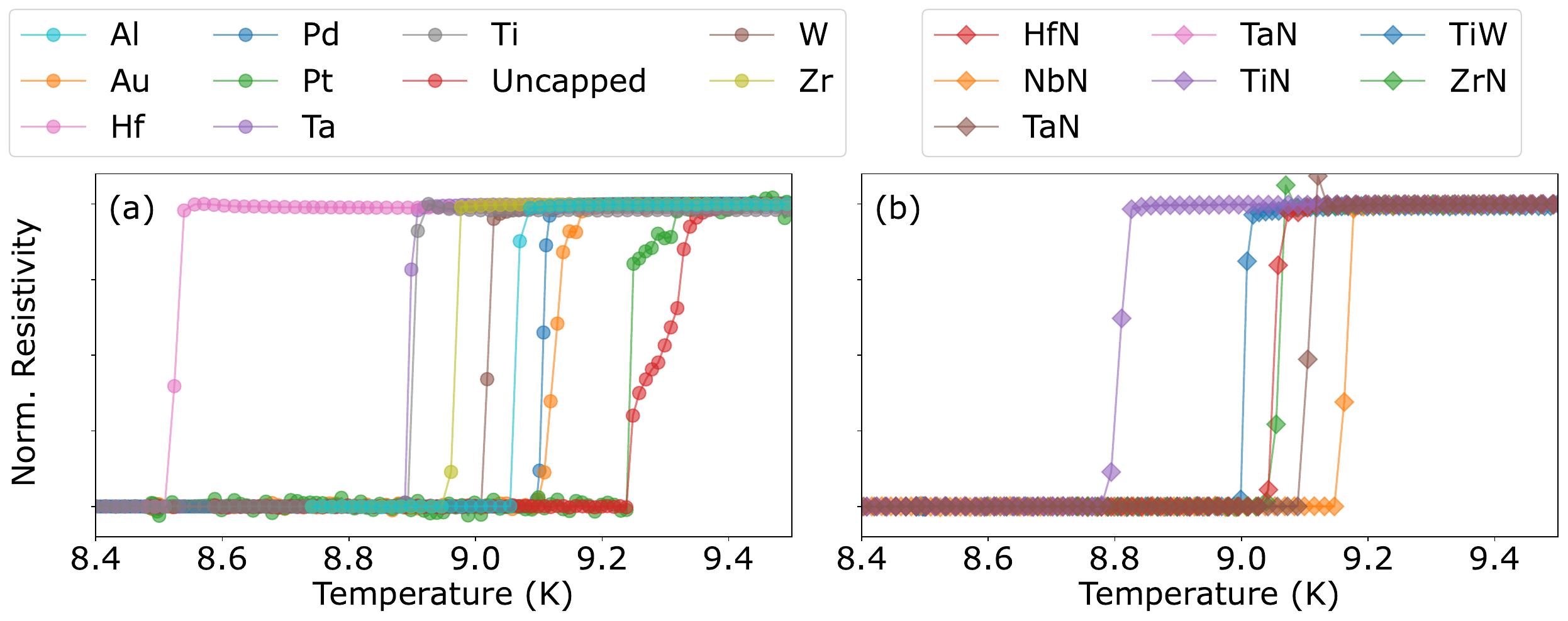}
    \caption{Critical temperature graphs for uncapped and capped Nb samples. 
    (a) Films capped with metals. 
    (b) Films capped with nitrides or intermetallics.}
    \label{fig:SI-Tc}
\end{figure*}

Superconducting critical temperature (T\textsubscript{c}) measurements are shown in Figure \ref{fig:SI-Tc}. 
Data is normalized to the resistivity at \SI{9.5}{\kelvin}.
T\textsubscript{c} data was not collected for the Mo or the Zr-Y capped samples.
T\textsubscript{c} measurements were taken in a Quantum Design PPMS DynaCool with a base temperature of \SI{1.8}{\kelvin}. 
We see that all samples have a T\textsubscript{c} greater than \SI{8.5}{\kelvin}.
Measurements were done un-patterned films with an in-line 4-pt measurement geometry.

\section{X-ray Photoelectron Spectroscopy}\label{app:XPS}

All XPS measurements were conducted in a ThermoScientific Nexsa G2 Surface Analysis system with a chamber pressure around $2\times$ 10$^{-7}$~\si{Torr}. 

The system uses a monochromated Al K-alpha X-ray source and the Fermi level is calibrated using a silver standard. 
Survey scans were collected using a \SI{400}{\micro\meter} spot size as counts per second (CPS) with a \SI{0.4}{\electronvolt} energy resolution. 
Fitting was done with a Shirley background in CasaXPS(v2.3.25)\cite{fairley_systematic_2021} using Thermo Scientific provided sensitivity factors calibrated for the Nexsa. 
Core level spectra were collected with a \SI{0.1}{\electronvolt} energy resolution. 

Nb and NbO are fit using asymmetric Voigt-like lineshapes, and NbO\textsubscript{2} and Nb\textsubscript{2}O\textsubscript{5} are fit using symmetric lineshapes.

\section{Annealing measurements}\label{app:anneal}

Samples were heated on a hot plate in the fume hood at \SI{200}{\celsius} for one hour, and vacuum annealed in a tube furnace with pressure below $5\times10^{-6}$~\si{Torr} at \SI{200}{\celsius} for \SI{1}{\hour}.
Samples were heated from \SI{25}{\celsius} with a ramp rate of \SI{10}{\celsius\per\min}, and then cooled for \SI{1}{\hour} after the anneal and removed at \SI{50}{\celsius}.

\begin{figure*}
    \centering
    \includegraphics[width=\linewidth]{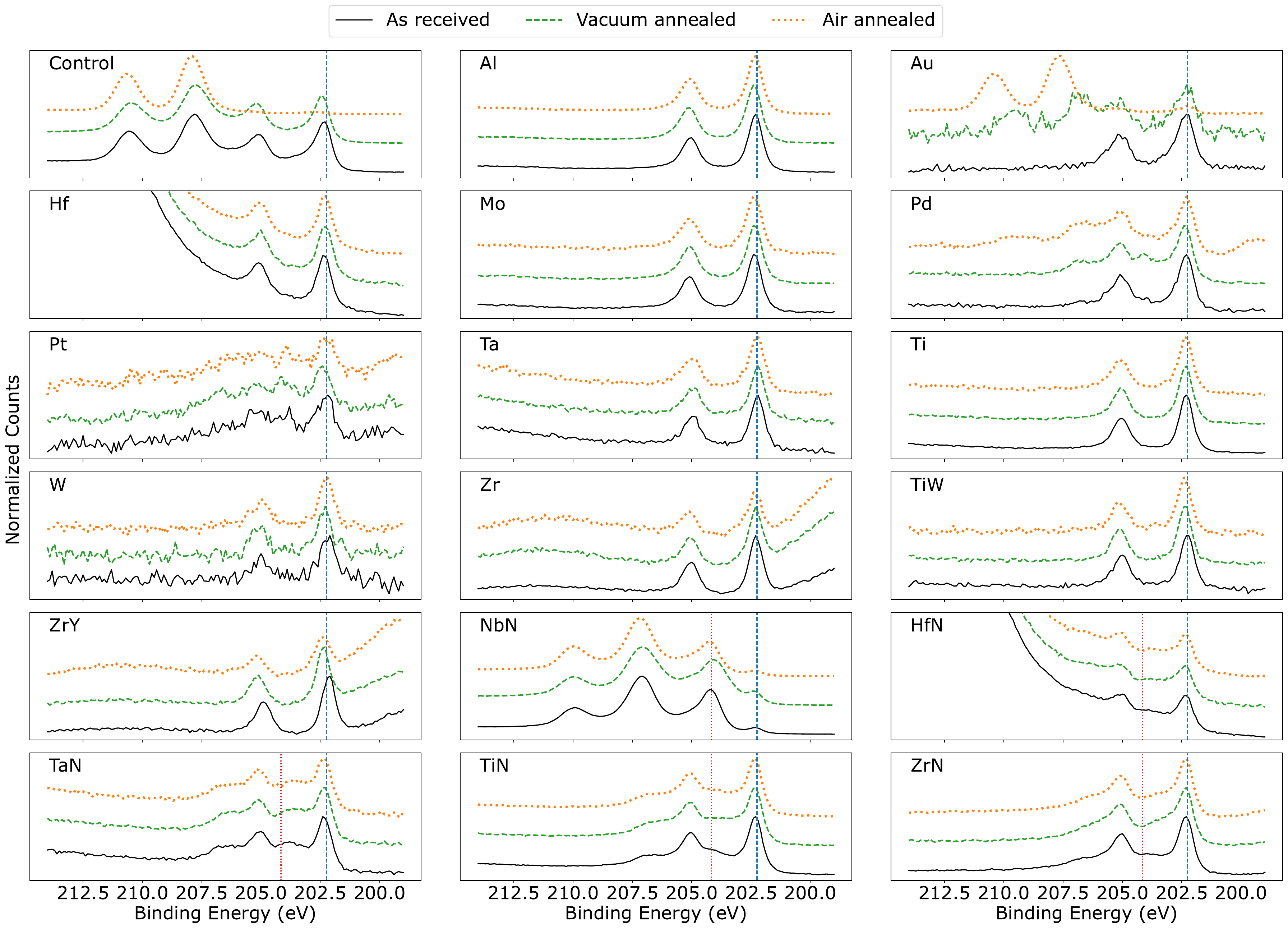}
    \caption{Compiled data of the Nb 3d core level scan of a control uncapped sample and various metal capped samples as received (black solid), after vacuum annealing (orange dashed) and after air annealing (green dotted). 
    Data is normalized to the Nb 3d\textsubscript{5/2} peak and thus cutoff at higher BE for Hf and HfN. 
    Data corresponds to the summary shown in Figure \ref{fig:annealing_table}. 
    For all samples, we see clear doublet peaks of the Nb metal, with the Nb 3d\textsubscript{5/2} peak around \SI{202.25}{\electronvolt} (blue dashed line). 
    For nitrides, we assume a peak at \SI{204.15}{\electronvolt} (red dotted line), according to data from the NbN sample.}
    \label{fig:SI-anneal}
\end{figure*}

See Figure~\ref{fig:SI-anneal} for the Nb 3d scan of all capping layers. 
The full set of scans of other peaks, including the survey, C 1s, O 1s, and respective metals, can be found in the respective VMS files in the repository.
All data was taken on the surface without in-situ etching in the XPS.
No binding energy correction was conducted.
Data is normalized to the Nb peak for all samples.
The low signal-to-noise ratio (SNR) can be explained by the difference in the mean free paths of electrons in the various caps. 
Furthermore, the various oxides of the cap metals formed increases the thickness of the cap commensurate with the oxidation of the cap.
Although the caps are nominally all \SI{5}{\nano\meter} thick, the differences in both rate and structure of the oxide growth result in different cap thicknesses, further contributing to differences in SNR.

The control sample shows that the vacuum annealing does not significantly affect the oxidation of the Nb, while the air annealing significantly increases the oxide.
Pd, Pt, Ta, W, Zr, TiW, and ZrY were later etched to confirm the conclusions drawn from the unetched data (see repository).

We use the NbN peak from this sample to reference the likely nitride peak location in the other nitride capped samples, which may not be fully accurate as the NbN\textsubscript{x} peak has a large standard deviation for its location, especially as it is likely non-stoichiometric~\cite{justin_gorham_nist_2012}.
It is evident that the TiN and ZrN caps have a peak forming near \SI{203}{\electronvolt} after annealing, which is likely a Nb suboxide.
It is possible for the peak to be a nitride, which is difficult to distinguish. 
However, the formation energy of the Nb nitride indicates that it is highly unlikely that nitride can be formed at a temperature as low as 200$^\circ$C, either under vacuum or in air~\cite{reed_free_1971}.
HfN and TaN have no such increase, and thus likely do not have any suboxide formation under annealing.
We note that the nitride peak in the TaN-capped sample does not line up with the peak in the NbN sample, but we attribute it to the variation in the NbN\textsubscript{x} peak.

\section{Chemical strip bath}\label{app:AZ}

\begin{figure*}
    \centering
    \includegraphics[width=\linewidth]{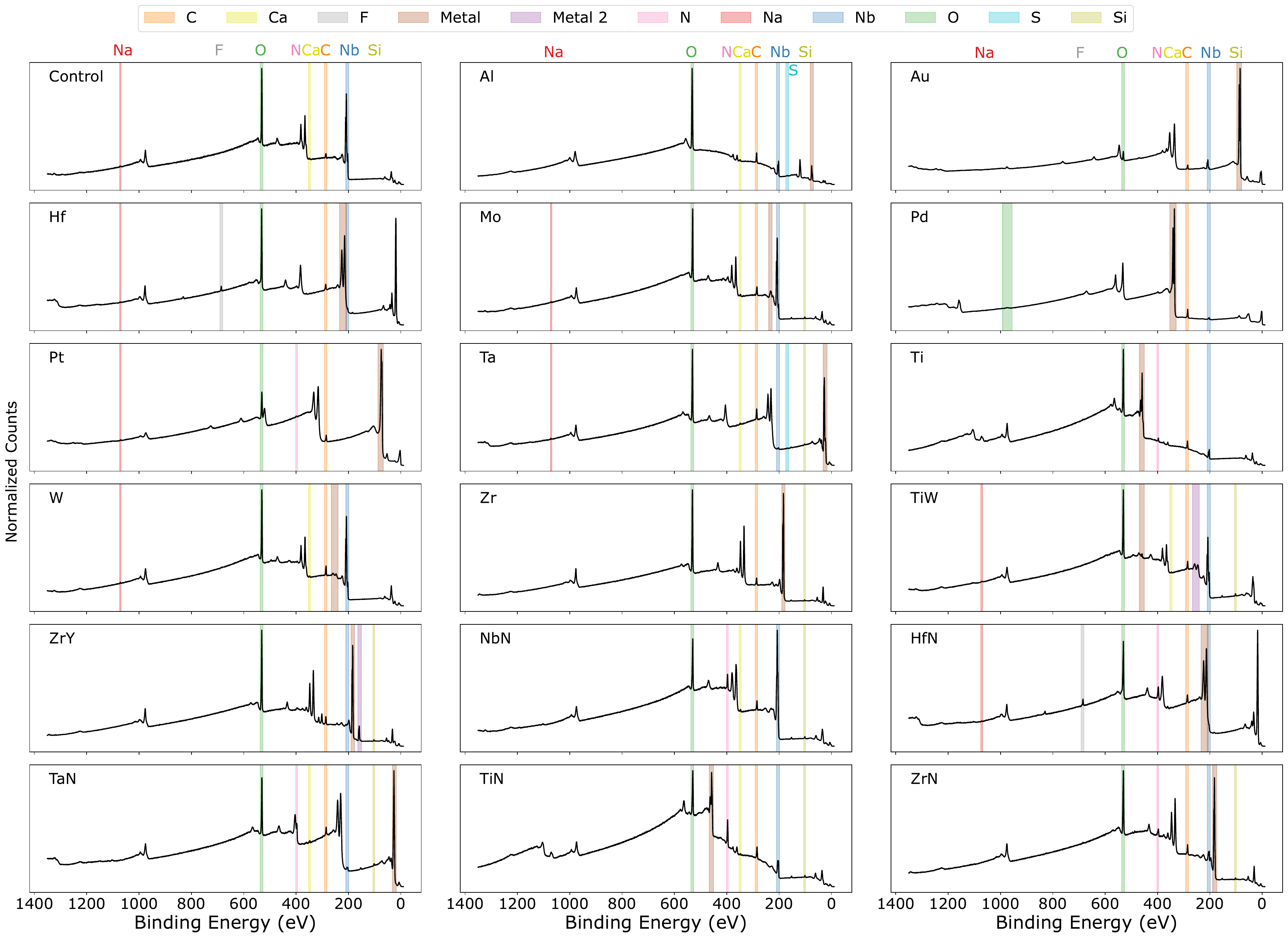}
    \caption{Compiled survey data of control Nb and various capped metal samples after AZ 300T strip bath. 
    Primary peaks are highlighted for each of the elements seen on the surface. 
    All other peaks are accounted for as additional peaks of the observed elements.}
    \label{fig:SI-AZ300T}
\end{figure*}

The survey scans post-AZ 300T strip bath are shown in figure \ref{fig:SI-AZ300T}. 
The relative atomic percentages are given in Table \ref{tab:SI-AZ300T}.

Prominent Nb peaks can be seen in the control sample, as well as the Mo, W, Ti-W, and NbN samples. 
While it is expected in the control and NbN sample, the prominent peak in the others indicate that the capping metal was significantly damaged by the AZ 300T process and has exposed the Nb layer which has then oxidized.
For the Pd sample, the Pd 3p peak overlaps with the O 1s peak, and therefore the O KL1 peak had to be used to determine atomic percentage of oxygen.
For the W peaks, the W 4f peak overlaps with the W 5p peak, and therefore we use the W 4d peak for atomic percentage. 

We see from Table \ref{tab:SI-AZ300T} that Na and Ca are common contaminants across a variety of samples, including the control.
As discussed before, this is likely from the specific chemical composition of the components in AZ 300T, or the cleanliness of the strip bath in general, which might leech common ions such as Na and Ca. 
Si is possibly from volatile silicates or Si dust from handling the samples
The F and S contamination are surprising, as there are no discernible sources of either during any of the processing. 
The samples were placed in the same bath with other samples in this table, and there seems to be no F or S elsewhere.

\begin{table*}[]
    \centering
    \begin{tabular}{c|c|c|c|c|c|c|c|c|c|c|c} \hline
Element	&C &O&Nb &Metal & Metal 2  &N&Na &Ca &F &S &Si \\ \hline \hline
Control&10.35\%&58.5\%&30.6\%&	-& -  &-&0.3\%&0.4\%&- &- &- \\ \hline
Al	&13.2\%&45.0\%& 2.5\%&38.5\%&	- &- & - &0.4\%&- &0.4\%&- \\ \hline
Au	&26.3\%&19.8\%& 9.3\%&44.6\%&	- &- & - &- &- &- &- \\ \hline
Hf	&13.7\%&58.1\%& 0.2\%&25.4\%&	- &- & 0.2\%&- &2.5\%&- &- \\ \hline
Mo*	&16.6\%&53.6\%& 23.7\%&2.5\%&	- &- & 0.3\%&0.6\%&- &- &2.8\%\\ \hline
Pd	&39.8\%&10.7\%& 1.1\%&48.5\%&	- &- & - &- &- &- &- \\ \hline
Pt	&21.4\%&37.3\%& -	&38.9\%&	- &1.9\%& 0.6\%&- &- &- &- \\ \hline
Ta	&19.6\%&52.2\%& 0.3\%&24.9\%&	- &- & 0.2\%&0.6\%&- &0.4\%&2.0\%\\ \hline
Ti	&15.4\%&48.2\%& 2.1\%&30.8\%&	- &3.6\%& - &- &- &- &- \\ \hline
W*	&19.6\%&52.9\%& 25.0\%&1.8\%&	- &- & 0.4\%&0.3\%&- &- &- \\ \hline
Zr	&13.6\%&58.6\%& -	&26.2\%&	- &- & - &- &- &- &1.7\%\\ \hline \hline
Ti-W*	&16.2\%&53.2\%& 21.4\%&0.6\%& 	4.5\% &- & 0.2\%&0.2\%&- &- &3.8\%\\ \hline
Zr-Y	&14.1\%&56.2\%& 0.1\%&22.3\%&	5.3\% &- & - &- &- &- &2.0\%\\ \hline \hline
NbN	&14.5\%&35.7\%& 29.5\%&-&	- &16.8\%& - &0.5\%&- &- &2.9\%\\ \hline
HfN	&16.9\%&40.4\%& 0.2\%&21.3\%&	- &17.4\%& 0.2\%&- &3.7\%&- &- \\ \hline
TaN	&17.3\%&39.9\%& 0.8\%&27.8\%&	- &11.2\%& - &0.7\%&- &- &2.3\%\\ \hline
TiN	&14.6\%&28.3\%& 4.0\%&25.8\%&	- &25.0\%& - &0.3\%&- &- &2.2\%\\ \hline
ZrN	&16.0\%&47.2\%& 2.5\%&23.8\%&	- &8.4\%& - &- &- &- &2.1\%\\ \hline

    \end{tabular}
    \caption{Relative atomic percentages of elements visible on the surface of capped Nb-metal bilayers after a chemical strip bath in AZ 300T. 
    Stars ($*$) indicate caps which were damaged by the AZ 300T.}
    \label{tab:SI-AZ300T}
\end{table*}

\section{Acid cleaning}\label{app:Acid}

\begin{figure*}
    \centering
    \includegraphics[width=\linewidth]{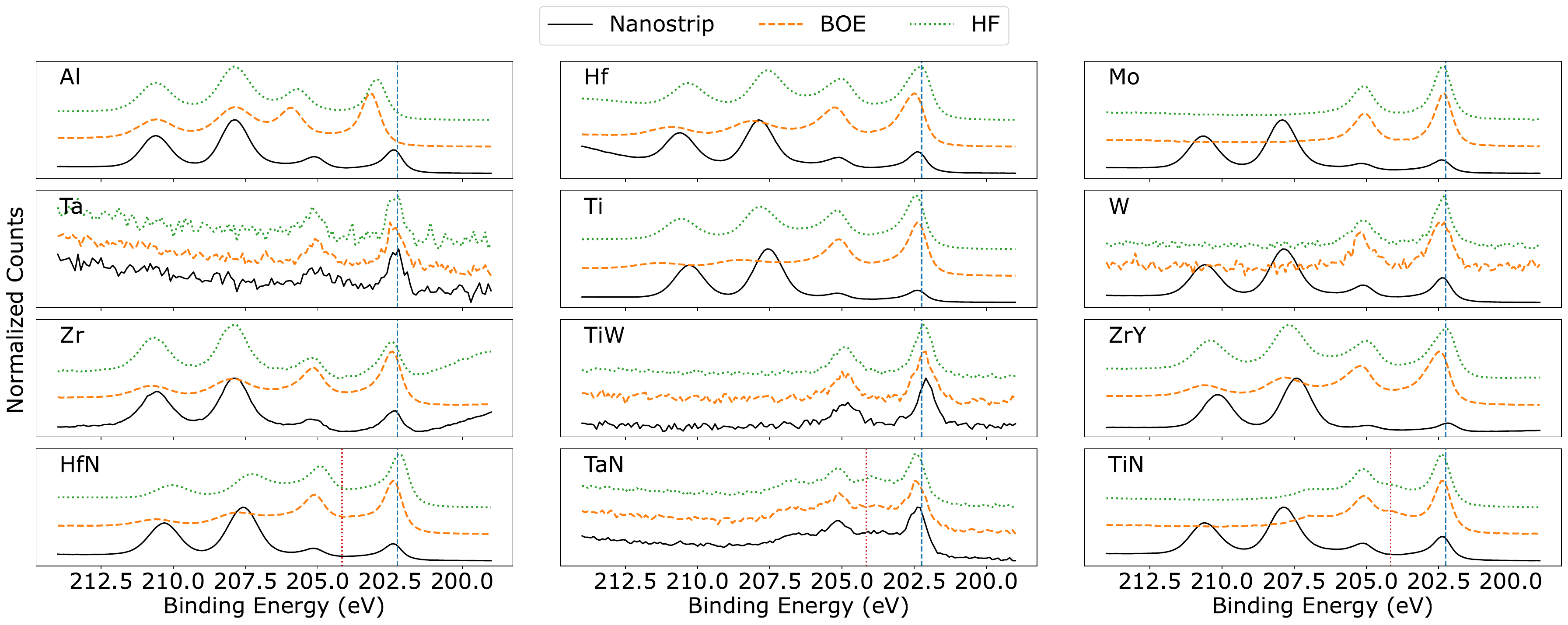}
    \caption{Compiled data of the Nb 3d core level scan of a control uncapped sample and select capped samples after three different acid cleans -- Nanostrip, 10:1 BOE, and 2\% HF. 
    For all samples, we see clear doublet peaks of the Nb metal, with the Nb 3d\textsubscript{5/2} peak around \SI{202.25}{\electronvolt} (blue dashed line). 
    For nitrides, we assume a peak at \SI{204.15}{\electronvolt} (red dotted line), according to data from the NbN sample. }
    \label{fig:SI-acid}
\end{figure*}

Figure \ref{fig:SI-acid} shows the Nb 3d core level spectra of all capped metal films after acid cleaning processes in Nanostrip, BOE, and HF. 
Further data of the full scans of the respective cap metal spectra, the survey scans, and the carbon and oxygen spectra can be found in the repository. 

We see that a majority of the samples have significant oxidation of the niobium, indicating that the caps were etched and/or damaged by the respective acids. 
Al, HF, Ti, Zr, ZrY, and HfN were damaged by all three acid processes, significantly eroding the cap metal and etching the underlying niobium. 
The high SNR indicates the lack of any overlayer affecting the mean free path of the electrons. 
For several of these, there is no indication of any cap metals remaining on the surface.

On the other hand, Mo, W, and TiN are only damaged by the Nanostrip, and show no damage to the Nb layer after either BOE or HF processing. 

TiW is an outlier, wherein the Ti has been etched away but the W still remains to some extent (in all three acid cleans). 
This is likely to be removed as well if we continue etching, although looking at the data from W, it may survive HF and BOE processing. 
Regardless, the etching is still problematic as it will introduce variations in the intermetallic composition, and thus should not be used as a cap. 

Ta and TaN show no change in the Nb spectra post-acid processing, and thus are candidates for optimization of acid cleaning post-processes for improved resonator performance. 

Table \ref{tab:SI-AcidXPS} shows the relative atomic percentages of the elements seen on the surface of the samples that survived the respective acid cleans. 
There are no Na, Ca, or Si signals on any of the samples.
This supports the hypothesis that the Na and Ca contamination are likely from the stripping process, and the Si, which may be from volatile silicates or dust from handling, is etched away during the process because the residues are likely particulate and mostly oxidized, and thus removed by the various acids. 

There is added Fluorine residues on the surfaces post-BOE and -HF cleans, which is a well-known phenomenon.
The samples were measured nominally the same day as the acid cleanings were done, and thus a high fluorine residue is expected, which is similar to what would occur if samples are loaded into the dilution fridge within a few hours of conducting an acid clean, as is the norm. 

It is interesting to note that the W-capped sample has no F-residue in post-HF (and minimal residue post-BOE). 
This could indicate the fluorine does not chemically adsorb onto the W surface well, and thus forms minimal residues. 
If halogenated surfaces are found to host increased TLS losses due to dangling bonds, W may be an element that could be considered for it's potential to minimize that avenue of loss. 
\begin{table*}
    \centering
    \begin{tabular}{c|c|c|c|c|c|c|c|c} \hline
 Acid&Cap&C &O&Nb &Metal  &N&  F&S\\ \hline \hline
\multirow{5}{*}{2\% HF}&Mo&27.1\%&35.7\%&2.7\%&	 34.5\%&-&   -&-\\ \cline{2-9}
 &Ta&27.4\%&49.0\%& -& 21.9\%&-&   1.8\%&-\\ \cline{2-9}
 &W&23.8\%&34.4\%& 0.4\%& 40.0\%&1.4\%&   -&-\\ \cline{2-9}
 &TaN&28.7\%&28.5\%& 0.9\%& 26.4\%&13.3\%&   2.2\%&-\\ \cline{2-9}
 &TiN&22.4\%&14.9\%& 5.9\%& 23.4\%&28.4\%&   5.1\%&-\\ \hline \hline
\multirow{6}{*}{10:1 BOE}&Mo&30.0\%&29.8\%& 2.9\%& 34\%&-&   3.4\%&-\\ \cline{2-9}
 &Ta&16.2\%&53.1\%& 0.1\%& 25.6\%&-&   5.0\%&-\\ \cline{2-9}
 &W&29.8\%&26.6\%& 0.4\%& 41.2\%&1.3\%&   0.7\%&-\\ \cline{2-9}
 &TaN&23.3\%&26.3\%& 0.8\%& 25.8\%&12.6\%&   11.2\%&-\\ \cline{2-9}
 &TiN&22.1\%&8.6\%& 5.8\%& 21.6\%&32.3\%&   9.5\%&-\\ \hline \hline
 \multirow{2}{*}{Nanostrip}&Ta&17.0\%&57.3\%& 0.1\%& 24.91\%&-&  -&0.7\%\\ \cline{2-9}
 &TaN&17.7\%&40.7\%& 0.9\%& 27.5\%&12.6\%&   -&0.6\%\\ \hline

    \end{tabular}
    \caption{Relative atomic percentages of elements visible on the surface of capped Nb-metal bilayers that survived respective acid cleans. }
    \label{tab:SI-AcidXPS}
\end{table*}

\section{Resonator measurements}\label{app:resonator}
\subsection{Resonator fabrication}

Resonators were fabricated as done in prior work~\cite{olszewski_low-loss_2025}.
We detail the identical process here for completeness.

\begin{figure}
    \centering
    \includegraphics[width=\linewidth]{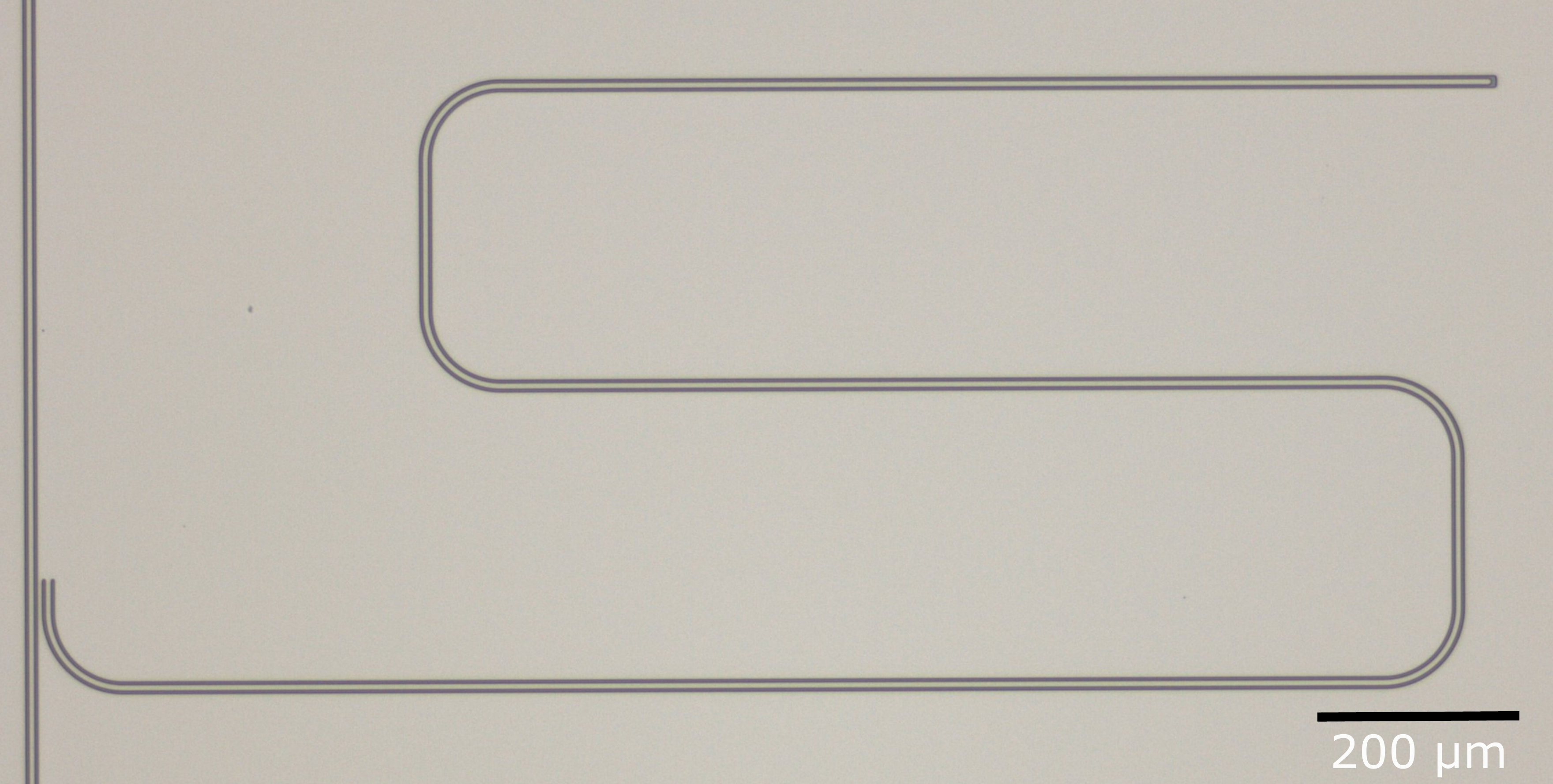}
    \caption{Optical microscope image of a CPW resonator made using the Zr-capped Nb film.}
    \label{fig:res_om}
\end{figure}

Wafers were prepared for deposition with an RCA clean, which is done by conducting SC-1 at \SI{70}{\celsius} (1:1:6 ammonium hydroxide, hydrogen peroxide and water), SC-2 at \SI{70}{\celsius} (1:1:6 hydrochloric acid, hydrogen peroxide and water) and spin-drying the wafer.
This was followed by dipping in 10:1 buffered oxide etch (BOE) for \SI{1}{\minute} followed by two baths in de-ionized water for \SI{30}{\second} each and drying with high-purity N\textsubscript{2} gas.
The respective films were then deposited on the prepared wafer.
After spinning MP S1813 photoresist at 3000 rpm on the wafers, the films are baked at \SI{90}{\celsius} for \SI{1}{\minute}. 
The resonators are patterned using a g-line stepper and developed in MP 321 MIF developer for 1 minute. 
The films are etched using a Pan ICP-RIE Plasma-Therm 770 etcher. 
There is a light etch using  BCl\textsubscript{3}, Cl\textsubscript{2}, and Ar gas in a 30:2:5~\si{sccm} ratio and RIE/ICP power of 26/800~\si{\watt} at 13~\si{mTorr}. 
Following this, the primary etch takes place, consisting of a the same BCl\textsubscript{3}/Cl\textsubscript{2}/Ar gas in a 20:30:5~\si{sccm} ratio with a RIE/ICP power of 12/800~\si{\watt} at 7~\si{mTorr}. 
The resist is stripped post-etch using heated AZ 300T resist stripper (AZ) at 80-\SI{90}{\celsius} for one hour, followed by sonications in isopropanol and deionized (DI) water, followed by rinsing in a second DI water and second isopropanol bath. 
The wafers are re-coated with S1813 resist for dicing, which is then stripped post-dicing using an identical stripping process to the one done post-etch.
Figure \ref{fig:res_om} shows an optical microscope image of one of the eight hangar-style CPW resonators in the Zr-capped Nb sample. 

\subsection{Fridge setup}

\begin{figure}
    \centering
    \includegraphics[width=\linewidth]{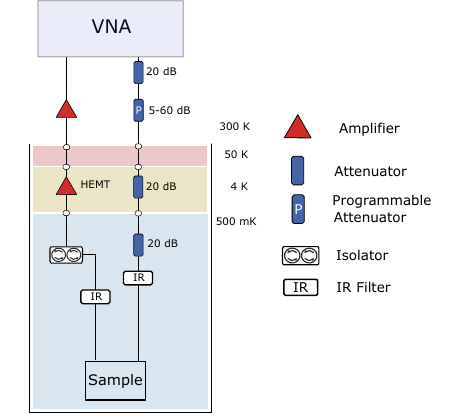}
    \caption{Fridge wiring diagram with a base temperature of \SI{500}{\milli\kelvin}.}
    \label{app:SD-wiring}
\end{figure}

The fridge used for the resonator measurements is a Bluefors small diameter refrigerator with a base temperature of \SI{500}{\milli\kelvin}. 
The fridge wiring diagram is shown in Figure \ref{app:SD-wiring}.
The input lines have about \SI{10}{dB} of attenuation from the lines themselves and \SI{60}{dB} attenuation from the attenuators.
The output line has an dual junction isolator, a HEMT (Low Noise Factory, \SI{37}{dB} gain), and a room temperature amplifier with about \SI{37}{dB} gain.
Both lines have eccosorb IR filters.
We use a Copper Mountain M5180 vector network analyzer (VNA) with \SI{10}{dB} to \SI{-50}{dB} dynamic range and a programmable attenuator with \SI{5}{dB} to \SI{55}{dB} tunability at room temperature.

\subsection{Measurement and analysis}
\label{app:resonator_measurements}
The resonator design has a \SI{3}{\micro\meter} gap between the center conductor and the ground plane and has a hangar resonator configuration, with eight $\lambda/4$ resonators at frequencies between 4 and \SI{8}{\giga\hertz} capacitively coupled to the feedline\cite{kopas_simple_2022}.

The complex transmission coefficient $S_{21}$ was measured with a VNA.
The $Q_{c}$ (coupled), $Q_{i}$ (internal), and $Q$ (loaded) quality factors were fitted using the diameter correction method (DCM)~\cite{mcrae_materials_2020,khalil_analysis_2012}:
\begin{equation}
    \label{eq:S21-fit}
    S_{21}(f) = 1-\frac{Q/\hat{Q_{c}}}{1+2iQ\frac{f-f_{0}}{f_{0}}},
\end{equation}
where $f$ is the frequency, $f_{0}$ is resonance frequency, and $\hat{Q_c}=Q_ce^{-i\phi}$ is the complex coupled quality factor.
The even distribution of measured points around the circle was achieved by using the homophasal point distribution (HPD) method~\cite{baity_circle_2024}.
The average photon number was calculated using~\cite{mcrae_materials_2020}:
\begin{equation}
    \langle n\rangle=\frac{E_{0}}{\hbar\omega}=\frac{2}{\hbar\omega_{0}^{2}}\frac{Z_{0}}{Z_{r}}\frac{Q^{2}}{Q_{c}}P_\mathrm{app}.
\end{equation}
$P_\mathrm{app}$ is the power applied to the device, and $\omega_{0}$ is the resonant frequency.
$Z_0$ is the characteristic impedance of the microwave package the resonator chip is connected to, with a designed value of $\SI{50}{\ohm}$, $Z_r$ is the characteristic impedance of each superconducting resonator, also with a designed value of $\SI{50}{\ohm}$.
To extract the full performance of the resonators, we measured $S_{21}$ in the range of $10$ to $10^{6}$ photons for $\langle n\rangle$.
The expected thermal photon population for a fully thermalized resonator is given by the Bose-Einstein distribution,

\begin{equation}
    n(f_{\rm CF},T) = \frac{1}{e^{hf_{\rm CF}/k_{B} T}-1},
\end{equation}
where $h$ is the Planck's constant, $f_{\rm CF}$ is the center frequency of the resonator, $k_{B}$ is the Boltzmann constant, and $T$ is the temperature.
For a \SI{6}{\giga\hertz} frequency resonator at \SI{500}{\milli\kelvin}, $\langle n \rangle \approx 1.28$ ($2.13$ for $f=4$~\si{\giga\hertz}).
To take into account thermal photons from higher stages, we use the equation~\cite{krinner_engineering_2019}
\begin{equation}
    n_i(f_{\rm CF}) = \frac{n_{i-1}(f_{\rm CF})}{A}+\frac{A_i-1}{A_i}n(f_{\rm CF},T),
\end{equation}
where $A = 20$ dB $=100$ is the attenuation at the given stage $i$. 
Conducting the cascading calculation, we find that $\langle n \rangle \approx 1.52$ ($2.48$ for $f=4$~\si{\giga\hertz}) 
Therefore, measuring the resonators below $1$ photon will not yield meaningful results.
Furthermore, the measurement between $1-10$ photons is likely not fully accurate due to the thermal photons.
Therefore, we use the data in the $10-100$ photons range for the calculation of $\delta_{\rm MP}$, as shown by the saturation in the data Fig.~\ref{fig:resonators}(b).

We define loss $\delta_{i}$ as the inverse of the quality factor $Q_{i}$:
\begin{equation}
    \delta_{i}\approx\tan\delta_{i}=\frac{1}{Q_{i}}.
\end{equation}
HP losses are the average losses above $10^{5}$ photons, and MP losses are average losses below $100$ photons.

At \SI{500}{\milli\kelvin} temperature of the cryostat, we do not anticipate a significant effect due to thermal quasiparticles (QPs) because of the high T\textsubscript{c} of the films.
In Olszewski et al.~\cite{olszewski_low-loss_2025}, the temperature dependence of the Nb resonator shows negligible change in internal loss below \SI{1}{\kelvin}.
In the limit $k_bT\ll \Delta$ and $\hbar \omega \ll \Delta$, the effect of QPs on the internal loss of the resonators can be estimated by the equation~\cite{crowley_disentangling_2023,fischer_nonequilibrium_2023,zmuidzinas_superconducting_2012,gao_physics_2008}
\begin{equation}
    \delta_{\rm QP}(T) = \frac{4\alpha}{\pi}\sinh \left(\frac{\hbar\omega}{2k_b T}\right) K_0\left(\frac{\hbar\omega}{2k_b T}\right)e^{-\frac{\Delta}{k_b T}},
\end{equation}
where $K_0$ is the zero-th order Bessel function of the second kind, $\Delta$ is the superconducting gap, and $\alpha$ is the kinetic inductance fraction ($\alpha<1$).
The calculated $\delta_{\rm QP}$ is on the order of $2.5\times10^{-16}\alpha$ for a Nb resonator at \SI{500}{\milli\kelvin} at \SI{6}{\giga\hertz}, and thus the effect of QP loss should be negligible for all samples, despite having slightly altered T\textsubscript{c} and $\Delta$ due to the capping layers.

\subsection{Resonator XPS}
\begin{figure}
    \centering
    \includegraphics[width=\linewidth]{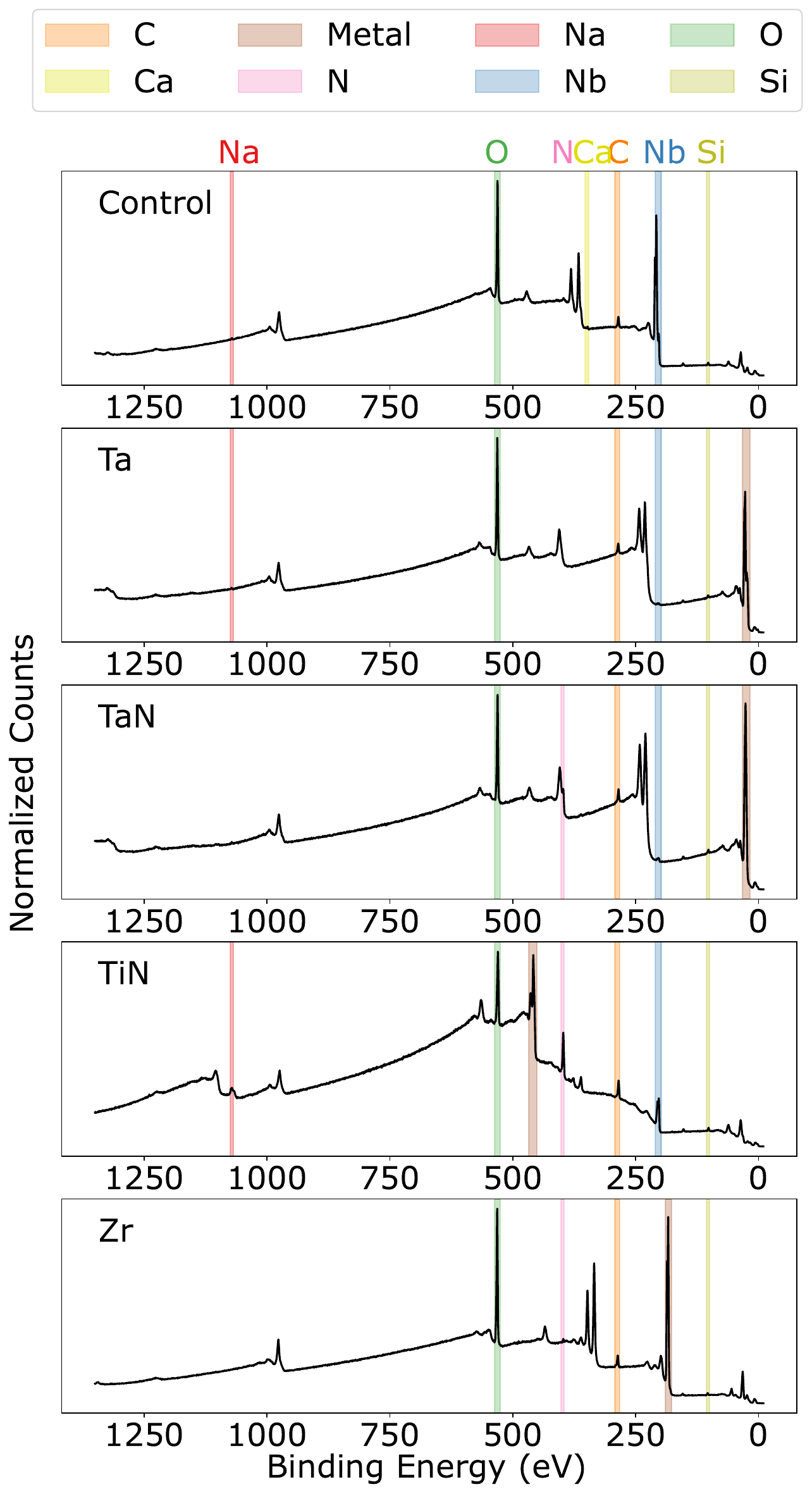}
    \caption{Survey data of control Nb and capped metal resonators after full fabrication. 
    Primary peaks are highlighted for each of the elements seen on the surface, while other peaks are marked as additional observed elements.}
    \label{fig:SI-ResXPS}
\end{figure}

To ensure consistency, one sample from each set of bilayers was measured in the XPS.
Survey spectra are shown in Fig.~\ref{fig:SI-ResXPS} and the relative atomic percentage are given in Tab.~\ref{tab:SI-ResXPS}.

All resonators show comparable surface contamination similar to the previous measurements (Table \ref{tab:AZ 300T}), however the control has the lowest carbon contamination.
The Si contamination is similar in all films, likely from the either volatile silicates being deposited on the surface or silicon dust from handling the chips.

\begin{table}
    \centering
    \begin{tabular}{c|c|c|c|c|c|c|c|c} \hline
Cap&C &O&Nb &Metal &N&Na &Ca &Si \\ \hline \hline
Control&12.7\%&56.2\%&27.6\%&	-&-&0.3\%&0.4\%&2.9\%\\ \hline
Ta	&14.4\%&55.5\%& 0.2\%&26.7\%&- & 0.2\%&-&2.9\%\\ \hline
Zr	&14.6\%&58.9\%& 0.4\%&23.2\%&0.9\%& - &- &2.1\%\\ \hline
TaN	&15.2\%&45.3\%& 0.5\%&27.6\%&8.3\%& - &-&3.0\%\\ \hline
TiN	&14.8\%&27.6\%& 4.3\%&25.4\%&24.9\%& 0.6\%&-&2.5\%\\ \hline

    \end{tabular}
    \caption{Relative atomic percentages of elements on the surface of capped Nb-metal bilayer resonators after fabrication including an AZ300T strip bath. }
    \label{tab:SI-ResXPS}
\end{table}

\clearpage
\bibliography{XPSBilayers.bib}

\end{document}